\documentclass[11pt,a4paper]{article}

\newcommand{\be}{\begin{eqnarray}}
\newcommand{\ee}{\end{eqnarray}}

\begin{document}

\title{\bf 
Condensation of charged bosons in plasma physics and cosmology}
\author{A.D.Dolgov$^a$\footnote{{\bf e-mail}: dolgov@fe.infn.it}\\
$^a$ \small{\em University of Ferrara  and INFN} \\
\small{\em Ferrara, FE44100, Italy }\\
\small{\em ITEP, Moscow, 117218, Russia}\\
}
\date{}
\maketitle

\begin{abstract}
The screening of impurities in plasma with Bose-Einstein condensate 
of electrically charged bosons is considered. It is shown that the 
screened potential  is drastically different from the usual Debye
one. The polarization operator of photons in plasma acquires infrared 
singular terms at small photon momentum and the screened potential 
drops down as a power of distance and even has an oscillating behavior, 
similar to the Friedel oscillations in plasma with degenerate fermions. 
The magnetic properties of the cosmological plasma with condensed 
W-bosons are also discussed. It is shown that W-bosons condense in the
ferromagnetic state. It could lead to spontaneous magnetization of the
primeval plasma. The created magnetic fields may seed galactic and
intergalactic magnetic fields observed in the present-day universe.
\end{abstract}

\section{Introduction}\label{s-intro}
The potential created by an electric charge  in plasma is
usually described by the well known Debye 
formula, see e.g. refs.~\cite{landau-9,kapusta}. The long
range Coulomb potential transforms into the exponentially decreasing 
Yukawa one:
\begin{equation}
U(r) = \frac{Q}{4 \pi r}\,\rightarrow \frac{Q\,\exp (-m_D r)}{4\pi r} \,,
\label{debye}
\end{equation}
because the time-time component of the photon propagator acquires 
a constant term which does not vanish when the three dimensional
photon momentum goes to zero. Instead of the vacuum $k^2$-term the 
inverse propagator becomes:  
\be 
k^2 \rightarrow k^2 +\Pi_{00} (k) =k^2 + m_D^2 \,,
\label{mD}
\ee
where e.g. for relativistic fermions\cite{fradkin}: 
\be 
m^2_D = e^2\left( {T^2}/{3} + {\mu^2}/{\pi^2} \right).
\label{mD2-rel}
\ee
These results are true if the fermions in plasma are not
strongly degenerate and if the charged bosons do not condense. 
The modification of the Debye screening in the case of 
degenerate fermions was studied half a century ago~\cite{friedel}, but the impact of
Bose-Einstein condensate (BEC) on the screening of impurities was considered
only very recently~\cite{gaba,dlp}. 
Surprisingly in the presence of BEC the screened
potential drastically changes and becomes even an oscillating function
of distance, which decreases either exponentially or as a power of $r$.
 
In the first part of this talk I discuss the impact of BEC of a charged
scalar field on the screening of charged impurities in plasma.
It is based on the works done in collaboration with A. Lepidi
and G. Piccinelli~\cite{dlp}. The second part about ferromagnetism
of the condensate of charged vector bosons is based on 
paper~\cite{dlp-vec} of the same group. More detailed list of
references can be found in these papers.

Let us consider electrically neutral plasma with large electric
charge density of bosons which is compensated by charged fermions.
Bosons condense when their chemical potential reaches the
maximum allowed value:
\be 
\mu_B = m_B\,.
\label{mu-B-max}
\ee
This can be easily seen from the consideration of the kinetic equation.
Indeed the equilibrium distribution of bosons, if and only if their chemical
potential is equal to their mass, takes the form:
\be 
f_B^{(eq)}  = C \delta^{(3)} ({\bf q}) +
\frac{1}{\exp\left[(E-m_B)/T\right] - 1}\,,
\label{fB-eq}
\ee
where the constant $C$ is the amplitude of the condensate. One can
check that $f_B^{(eq)}$ annihilates the collision integral for an
arbitrary C. Thus, by definition, it is an equilibrium distribution.
It is worth noting that the equilibrium distributions are
always functions of two parameters: temperature, $T$, and chemical
potential, $\mu$, when $\mu<m$, and temperature $T$, and the amplitude
of the condensate, $C$, when $\mu = m$. 

We calculate the time-time component of the photon polarization
operator in a simple straightforward way perturbatively solving operator
equation of motion for the electromagnetic field (Maxwell equations)
and averaging them over medium. One can reach the goal without
applying to more refined real or imaginary time
methods (for a review of these methods see
e.g. Ref.~\cite{kapusta}).

\section{Maxwell equations in thermal bath}\label{QED-med}
The standard Lagrangian of QED with interacting electromagnetic field 
and charged scalar and fermion fields 
with masses $m_B$ and $m_F$ respectively and with opposite electric charges $\pm e$ has the form:
	\begin{eqnarray}
		\mathcal{L} = -\frac{1}{4} F_{\mu\nu} F^{\mu\nu} - m_B^2 |\phi|^2
	+ |(\partial_{\mu} + i\,e A_{\mu}) \phi |^2
	+ \bar \psi (i \partial \!\!\! / - e A \!\!\! /  - m_F) \psi.
\label{L-qed}
	\end{eqnarray} 

The equations of motion for the electromagnetic
and charged scalar and spinor fields are respectively: 
	\begin{eqnarray}
		 (i \partial \!\!\! /  - m) \psi (x) = e A \!\!\! / \psi (x)\,,
\label{psi-eq-of-mot}	
	\end{eqnarray}
	\begin{eqnarray}
	\label{Phi_EOM}
	(\partial_{\mu} \partial^ {\mu} + m^2) \phi (x) &=& \mathcal{J}_\phi (x)\,,
		\end{eqnarray}
	\begin{eqnarray}
	\label{phot_EOM}
	\partial_{\nu} F^{\mu\nu} (x) &=& \mathcal{J}^\mu (x)\,,
	\end{eqnarray}
where the currents $\mathcal{J}$ can be written as: 
	\begin{eqnarray}
	\label{phi_current}
	\mathcal{J}_\phi (x) &=&
	-  i\,e \bigg[  \partial_{\mu} A^{\mu} (x) + 2 A_{\mu} (x) \partial^{\mu} \bigg] \phi (x)
	+ e^2 A^{\mu} (x) A_{\mu} (x) \phi (x)\,,
	\end{eqnarray}
	\begin{eqnarray}
	\label{EOM_current}
	\mathcal{J}^\mu (x)
	&=& - i\,e \bigg[(\phi^{\dag} (x)\partial^{\mu} \phi(x) ) 
- (\partial^{\mu} \phi^{\dag}(x)) \phi(x) \bigg] \nonumber\\
	&+& 2 e^2 A^{\mu} (x)|\phi(x)|^2 
	- e \bar \psi (x) \gamma^\mu \psi (x).
	\end{eqnarray}
Here $F_{\mu\nu} = \partial_\mu A_\nu - \partial_\nu A_\mu$ and
$\mathcal{J}^\mu$ (\ref{EOM_current}) is the total electromagnetic current 
of bosons and fermions.

Operator equations (\ref{psi-eq-of-mot}) and (\ref{Phi_EOM}) can be
formally solved as:
	\begin{eqnarray}
	\label{field_1_ord_exp}
	\phi(x) &=& \phi_0 (x) + \int d^4y \, G_B(x-y)
        \mathcal{J}_{\phi} (y) \,, \\
	\psi(x) &=& \psi_0 (x) + \int d^4y \, G_F(x-y) e A\!\!\! / (y) \psi(y)\,,
\label{psi-of-x}
	\end{eqnarray}
where $G_B$ and $G_F$ are the Green's functions of bosons and ferminos 
respectively and the zeroth order fields satisfy the free equations of motion:
	\begin{eqnarray}
	\label{phi_0_ord}
	(\partial_{\mu} \partial^ {\mu} + m_B^2) \phi_0 (x) = 0 ,
	\hspace{1cm} 
	( i \partial \!\!\! / - m_F ) \psi_0 (x) = 0 
	\end{eqnarray}
and are quantized in the usual way:
\begin{eqnarray}
	\label{phi-0-x}
	\phi_0 (x) &=& \int \frac{d^3q}{\sqrt{(2\pi)^3 2 E}}
	\left[ a({\bf q}) e^{-i q x} + b^\dag ({\bf q}) e^{iqx} \right]\,,\\
	\psi_0 (x) &=& 
	\int \frac{d^3q}{\sqrt{(2\pi)^3}} \sqrt{\frac{m_F}{E}}
	\left[ c_r({\bf q}) u_r({\bf q}) e^{-i q x} + d_r^\dag 
({\bf q}) v_r ({\bf q}) e^{iqx}
	\right].
\label{psi-0}
	\end{eqnarray}

Substituting solutions (\ref{field_1_ord_exp},\ref{psi-of-x})
into eq.~(\ref{phot_EOM})
we obtain the Maxwell equations with the lowest order
corrections to the electromagnetic current:
	\begin{eqnarray}
	\label{partial_drv_Fmunu}
	\partial_\nu F^{\mu\nu} (x) &=&
	-i\,e \bigg[(\phi_0^{\dag} (x)\partial^{\mu} \phi_0(x) ) 
	- (\partial^{\mu} \phi_0^{\dag}(x)) \phi_0(x) \bigg]
	- e \bar \psi_0 (x) \gamma^\mu \psi_0 (x) 
	\cr\cr
	&-& i e \, \phi_0^\dag (x) \partial^{\mu} \left[ \int d^4y \, G_B(x-y) 
	\mathcal{J}_{\phi_0} (y) \right]
	- i e \left[ \int d^4y \, G_B(x-y) \mathcal{J}_{\phi_0} (y) \right]^\dag 
	\partial^{\mu}  \phi_0 (x)
	\cr\cr 
	&+& i e \, \partial^{\mu} \phi_0^{\dag}(x) \left[ \int d^4y \, G_B(x-y) 
	\mathcal{J}_{\phi_0} (y) \right]
	+ i e \, \partial^{\mu} \left[  \int d^4y \, G_B(x-y) \mathcal{J}_{\phi_0} (y) \right] ^\dag
	\phi_0(x)
	\cr\cr
	&-& e \bar \psi_0 (x) \gamma^\mu \int d^4y \, G_F(x-y) e A\!\!\! / (y) \psi(y)
	- e  \left[ \int d^4y \, \bar \psi_0 (y) A\!\!\! / (y) \, G_F^*(x-y) \right] \gamma^\mu \psi_0(x)
	\cr\cr
	&+& 2 e^2 A^{\mu} (x)|\phi_0(x)|^2 .
	\end{eqnarray}
To derive the Maxwell equations with the account of the impact of medium on 
the photon propagator 
we have to average operators $\phi$ and $\psi$ over the medium. 
The first term in eq. (\ref{partial_drv_Fmunu}), linear in $e$, is non-zero if 
the medium is either 
electrically charged or possesses an electric current.
We assume here that this is not the case, i.e.
the medium is electrically neutral and ``current-less''.

The products of creation-annihilation operators averaged over the medium have the 
standard form:
	\begin{eqnarray}
	\label{thermal_averages}
	\langle a^\dag({\bf q}) a({\bf q}') \rangle =
f_B (E_q) \delta^{(3)} ({\bf q} - {\bf q}'),\\
	\langle a({\bf q}) a^\dag({\bf q}') \rangle = [1 + f_B (E_p)] \delta^{(3)} 
({\bf q} - {\bf q}'),\\
	\langle c^\dag({\bf q}) c({\bf q}') \rangle = f_F (E_p) \delta^{(3)} 
({\bf q} - {\bf q}'),\\
	\langle c({\bf q}) c^\dag({\bf q}') \rangle = [1 - f_F (E_p)] \delta^{(3)} 
({\bf q} - {\bf q}'),
	\end{eqnarray}
where $f_{F,B}(E_q)$ is the energy dependent fermion/boson distribution function, 
which may be 
arbitrary since we assumed only that the medium is homogeneous and isotropic.
We also assumed, as it is usually done, that the non-diagonal matrix elements of
creation-annihilation operators vanish on the average due to decoherence.
For the vacuum case $f_{F,B}(E) = 0$ and we obtain the usual vacuum average values of 
$\langle a a^\dag \rangle$ and $\langle a^\dag a\rangle = 0$, which
from now on will be neglected because we are 
interested only in the matter effects.
As a result we obtain linear but non-local equation for electromagnetic field $A_\mu (x)$,
for which it is convenient to perform the Fourier transform:
\begin{eqnarray}
	\label{A-mu-k}
	A^\mu (k) = \int \frac{d^4 x}{(2\pi)^3} e^{-ikx} A^{\mu} (x). 
	\end{eqnarray}
Finally we find that field $A^\mu (k)$ satisfies the equation
	\begin{eqnarray}
	\label{Phot_EOM_Pi_munu}
	\left[ k^\rho k_\rho g^{\mu\nu} - k^\mu k^\nu + \Pi^{\mu\nu} (k)\right] A_\nu (k) 
	= \mathcal{J}^\mu (k),
	\end{eqnarray}
which is equivalent to the photon equation of motion (\ref{partial_drv_Fmunu})
in momentum space.

In this way the photon polarization tensor, which contains
contributions from the charged bosons
and fermions, $\Pi_{\mu\nu} (k) =\Pi_{\mu\nu}^{B} (k) + \Pi_{\mu\nu}^{F} (k)$,
according to eq. (\ref{partial_drv_Fmunu}),
can explicitly found in the lowest order in $e^2$: 
	\begin{eqnarray}
	\label{phot_pol_tensor_bos}
	\Pi_{\mu\nu}^{B} (k) 
	= e^2 \hspace{-0.1cm} \int \frac{d^3q}{(2 \pi)^3 E} 
	\left[ f_B (E) + \bar f_B (E) \right] 
	\left[  \frac{1}{2} \, \frac{(2q - k)_\mu (2q - k)_\nu}{(q
            -k)^2 - m_B^2} \right. \nonumber \\
\left.	+\frac{1}{2} \, \frac{(2q + k)_\mu (2q + k)_\nu}{(q + k)^2 - m_B^2} 
 -   g_{\mu\nu}\right]\,,
	\end{eqnarray}	
	\begin{eqnarray}
	\label{phot_pol_tensor_fer}
	\Pi_{\mu\nu}^{F} (k) &=&
2 e^2 \int \frac{d^3q}{(2 \pi)^3 E} 
	\left[ f_F (E) + \bar f_F (E) \right] 
\left[
	\frac{q_\nu (k+q)_\mu - q^\rho k_\rho g_{\mu\nu} + q_\mu
          (k+q)_\nu}{(k+q)^2-m_F^2}  \right. \nonumber \\
	&+& \left. \frac{q_\nu (q-k)_\mu + q^\rho k_\rho g_{\mu\nu} + q_\mu (q-k)_\nu}{(k-q)^2-m_F^2}
	\right]\,.
	\end{eqnarray}
The static properties of the medium are determined by the
time-time component of the polarization tensor in the 
limit of $\omega = 0$, which can be easily calculated from the above
expressions:
\begin{eqnarray}
\Pi_{00}^B (k) = - \frac{e^2}{2\pi^2}\,\int_0^\infty\,\frac{dq q^2}{E}
(f_B +\bar f_B)\left( 1 +\frac{E^2}{k q}\,\ln\bigg|\frac{2q +k}{2q -k}\bigg| \right),
\label{Pi-bos-int} \\
\Pi_{00}^F (k) = - \frac{e^2}{\pi^2}\,\int_0^\infty\,\frac{dq q^2}{E}
(f_F +\bar f_F)\left( 1 +\frac{E^2}{k q}\,\ln\bigg|\frac{2q +k}{2q -k}\bigg| \right).
\label{Pi-ferm-int} 
\end{eqnarray}
Here and in what follows $k$ and $q$ are respectively the absolute values of the spatial 
component of the photon and the charged particle momenta.
Expressions (\ref{Pi-bos-int},\ref{Pi-ferm-int}) coincide with the
well known ones found by other methods. Our new results for
screening come from an addition of the condensate term to $f_B$, eq. (\ref{fB-eq}).

After straightforward calculations we find that the time-time
component of the charge boson contribution into the photon
polarization tensor at zero frequency and small $k$ (but high $T$) 
has the form:
\begin{eqnarray}
 \Pi_{00}=[k^2 + e^2(m_0^2 + m_1^3/k + m_2^4/ k^2)] \,,
\label{max-eq}
\end{eqnarray}
where 
\begin{eqnarray}
m_0^2 &=& {2T^2}/{3} + C/{[(2\pi)^3m_B]} \nonumber\\
m_1^3 &=&  {m_B^2 T}/{2} \nonumber \\
m_2^4 &=& {4Cm_B}/{(2\pi)^3}  .
\label{m-0-1-2}
\end{eqnarray}
In fact the same dependence on $k$ is true for any temperature but the
coefficients $m_j$ may be different.

\section{Screening of electric charge}\label{scr-pot}
The screened potential is determined by the Fourier transformation of the
photon propagator $(k^2 - \Pi_{00} )^{-1}$:
\begin{eqnarray}
U(r) = Q \int \frac{d^3 k}{(2\pi)^3} \frac{\exp (i {\bf k r)}}{k^2 - \Pi_{00} (k)} =
\frac{qQ}{2\pi^2}\int_0^\infty \frac{dk k^2}{k^2 - \Pi_{00} (k)}\,\frac{\sin kr}{kr}.
\label{U-of-r}
\end{eqnarray}
If $\Pi_{00}$ is an even function of $k$, as is usually the case, the integration over
$k$ can be extended to the interval from $-\infty$ to $+\infty$ and the integral can be
taken as a sum over residues of the integrand. For example, if the term proportional to
$1/k$ can be neglected (low temperature case), 
$\Pi_{00}$ is evidently even and its poles can be easily found:
\begin{eqnarray}
k_j^2 = -\frac{e^2 m_0^2}{2}\pm \sqrt{\left[\frac{e^4 m_0^4}{4} - e^2m_2^4 \right]}
\approx \pm i e m_2^2.
\label{k-j-2}
\end{eqnarray}
The last approximate equality is formally true in the limit of 
vanishingly small $e$.
 
Thus in the presence of the charged Bose condensate
the ``Debye'' poles acquire the non-zero real parts:
\begin{eqnarray}
k_j = \pm \sqrt{e} m_2 \exp (\pm i\pi/4) \equiv
k_j' + ik''_j,
\label{k-j}
\end{eqnarray}
Non-zero $k'$ leads to the oscillating behavior of the potential
\begin{eqnarray}
U(r)_j \sim Q\, \frac{\exp ( - \sqrt{e/2} m_2 r) \cos (\sqrt{e/2} m_2 r)}{r}.
\label{U-j}
\end{eqnarray}

If the term proportional to $1/k$ is present in $\Pi_{00}$,
the calculations of the potential are slightly more complicated. Now the integration
path in the complex $k$ plane cannot be extended to $-\infty$ but the integration 
should be done along the real axis from 0 to $\infty$, then along infinitely large 
quarter-circle, and along the imaginary axis from $\infty$ to 0. The result would
contain the usual contributions from the poles and other singularities
(see below) in the complex $k$-plane and the integral
over the imaginary axis. The former gives the usual exponentially decreasing
potential, while the latter gives a power law decrease:
\begin{eqnarray}
 U \sim Q\, m_1^3/(e^2 m_2^8 r^6).
\label{power-U}
\end{eqnarray}
Notice that the potential is inversely proportional to the electric
charge squared. This is because we consider asymptotical behavior
of the screened potential at large distance, $r$, when the parameter
$e r$ formally tends to infinity. 

There are some other singularities in the integrand of
eq. (\ref{U-of-r}), which arise from the pinching of the contour of
integration over $q$ in eq. (\ref{Pi-bos-int}) by the poles of the
distribution function, $f_B$~(\ref{fB-eq}),  and 
the branch point of the logarithm. The
induced singularities in the complex k-plane give rise to the
screening effects analogous to the Friedel oscillations~\cite{friedel}.
If the first ``pinch'' dominates, the screened potential is:
\be 
U_1 (r) = - \frac{32\pi Q}{e^2 m_B r^2} \frac{e^{-z}}
{\ln^2(2\sqrt{2}z) }\,\sin z\,,
\label{U1-large-r}
\ee
where ${ z = 2r\sqrt{2\pi T m_B}}$. Notice that
${U_1(r)}$ is inversely proportional  to ${e^2}$ and
formally vanishes at ${T \rightarrow 0}$, but remains finite if 
${\sqrt{T m_B} r \neq 0}$ .

If all pinches are comparable, the screened potential drops down as a
power of distance:
\be 
U(r) \approx -\frac{3 Q}{2 e^2 T^2 m_B^3 r^6 \ln^3 (\sqrt{8m_B T } r)}.
\label{U-B-fin}
\ee
More details can be found
in the second paper of Ref.~\cite{dlp}.

\section{Condensation of vector bosons}\label{vec-cond}
The condensation of the charged vector
bosons~\cite{dlp-vec} of the electroweak group
might take place in the early
universe if the cosmological lepton asymmetry was sufficiently high. 
Condensation of vector particles differs from that of the scalars due
to the additional degree of freedom, their spin states. Depending upon the
interactions between the spins, they can be either aligned or
anti-aligned. These states are called respectively ferromagnetic and 
anti-ferromagnetic ones, see e.g. Ref.~\cite{Pethick_book}.
We show that $W$-bosons of the minimal electroweak theory condense
in the ferromagnetic state and spontaneous magnetization of the
primeval plasma could generate strong primordial magnetic fields on
macroscopically large scales.

Recently somewhat similar problem of condensation of deuterium nuclei 
in astrophysics has been studied  in Ref.~\cite{gaba-vec}. 
The authors argue that the interaction between deuterium nuclei forces 
them into the lowest spin antiferromagnetic state. 

In the minimal standard electroweak model the spin-spin interaction of
$W$-bosons is determined by the interaction between their magnetic
moments and their contact quartic coupling. The former can be found
from the analogue of the Breit equation for vector
particles which leads to the spin-spin potential of the
form~\cite{dlp-vec}:
	\begin{eqnarray}
	\label{U_spin}
	U^{spin}_{em} (r)=  \frac{\alpha \rho^2}{m_W^2} 
	\left[ 
	\frac{\left(\mathbf{S}_1 \cdot \mathbf{S}_2 \right)}{r^3}
	- 3 \,
	\frac{\left(\mathbf{S}_1 \cdot \mathbf{r} \right)  
\left(\mathbf{S}_2
 \cdot \mathbf{r} \right)}{r^5}
	- \frac{8 \pi}{3} \left(\mathbf{S}_1 \cdot \mathbf{S}_2 \right) 
\delta^{(3)}(\mathbf{r})
	\right]\,,
	\end{eqnarray}
where $\alpha = e^2/4\pi \approx 1/137$ and $\rho$ is the ratio of the
real magnetic moment of $W$ to its value in the electroweak theory. 
Since the plasma is supposed
to be neutral, the Coulomb interaction between the condensed $W$ is
compensated by the charged leptons. 

To find the energy shift of a pair of $W$-bosons due to this interaction
we need to average potential (\ref{U_spin}) over the W-wave
function. The wave function of the condensate is supposed to be angle
independent S-wave state. Thus the energy shift induced by the 
magnetic spin-spin interaction is expressed through the integral of
potential (\ref{U_spin}) over space:
	\begin{eqnarray}
	\label{En_shift}
	\delta E = \int \frac{d^3r}{V}  \, U^{spin}_{em} (r) = 
	- \frac{2 \, e^2 \rho^2}{3 \, V m_W^2} \left(\mathbf{S}_1 \cdot \mathbf{S}_2 \right),
	\end{eqnarray}
where V is the normalization volume.

Since $S_{tot}^2 = (S_1+S_2)^2 = 4 + 2 S_1 S_2$, the average value of $S_1S_2$ is
equal to
	\begin{eqnarray}
	\label{S-tot}
	S_1 S_2 = S_{tot}^2/2 - 2\,.
	\end{eqnarray}
For $S_{tot} = 2$ this term is $S_1 S_2 = 1 >0$, while for $S_{tot} =0$ it
is $S_1 S_2 = -2 <0$. Thus, if the spin-spin interaction is dominated
by the  interactions between the magnetic moments of $W$ bosons,
the state with their maximum total spin is energetically
more favorable and $W$-bosons should condense in the
ferromagnetic state. Such an interaction would lead to the spontaneous
magnetization of the vector particles in the early universe.   

Another contribution to the spin-spin interactions of $W$ comes from their
quartic self-coupling:
\begin{eqnarray}
L_{4W}= \frac{e^2}{2\sin^2\theta_W}
\left[ W_\mu^\dagger W^{\mu \dagger} W_\nu W^\nu  -(W_\mu^\dagger W^\mu )^2 
 \right]=
\frac{e^2  \left({\bf W}^\dagger \times{\bf W}\right)^2 }{2\sin^2\theta_W}.
\label{spin-spin-W}
\end{eqnarray}
It is assumed here that $\partial_\mu W^\mu =0 $ and thus only the spatial 3-vector
${\bf W}$ is non-vanishing, while $W_t =0$.
The Fourier transform of this term with proper (nonrelativistic)
normalization leads to 
\begin{eqnarray}
U^{(spin)}_{4W} = \frac{e^2}{8 m_W^2 \sin^2 \theta_W} \left( {\bf S }_1 
{\bf S}_2 \right) \delta^{(3)} ({\bf r}).
\label{U-spin-W}
\end{eqnarray}
Thus the quartic self-coupling of $W$ contributes only to the
spin-spin interaction whose sign is antiferromagnetic.

In the minimal standard model the interaction between the magnetic moments
of $W$ dominates and the condensed $W$-bosons have ferromagnetic
behavior. However, in some modification of the standard model
antiferromagnetic behavior is possible and $W$ would condense in the
state with zero or microscopically small total spin. In this case classical
vector field of $W$-bosons could not be created, though they would still
make the Bose condensed state.

The exchange of $Z^0$-bosons may also contribute to spin-spin
interactions of $W$. It can be shown that for non-relativistic $Z$
this contribution vanishes~\cite{dlp-vec}. However, if the momentum
carried by the virtual $Z$ is non-negligible in comparison with its
mass, the contributions of $Z$ and photon exchanges are similar.

The long range interactions between the magnetic moments 
of $W$-bosons, in principle, can 
be screened by the plasma effects. However, in contrast to electric
interactions, which are Debye-like screened, magnetic interactions in
pure electrodynamics (or in any other Abelian theory) are known to
remain unscreened. On the other hand, in 
non-Abelian theories the screening may occur in higher orders of
perturbation theory due to the violent infrared singularities,
which make impossible perturbative calculations~\cite{linde-screen} .
The screening may potentially change the relative strength of the 
electromagnetic spin-spin coupling, which is affected by screening effects, with
respect to the local quartic, $W^4$-coupling which is not screened. 
However, in the broken phase the system is
reduced to the usual electrodynamics, where screening is absent and
$W$-bosons would condense in the ferromagnetic state. In the unbroken phase
of the electroweak theory the answer is not yet known. 

If the magnetic intraction between the spins of W-bosons is screened 
the potential describing the magnetic spin-spin interaction is related to
amplitude (\ref{U_spin}) with a modified photon propagator. So it can be
written as: 
\begin{eqnarray}
U^{(spin)}_{em} ({\bf r})= - \frac{e^2\rho^2}{m_W^2}
\int \frac{d^3 q}{(2\pi)^3} \frac{\exp \left(i {\bf  qr}\right)}{(q^2+ \Pi_{ss}(\mathbf{q}))}
	\left[ 
	q^2 \left(\mathbf{S}_1 \cdot \mathbf{S}_2 \right)
	- \left(\mathbf{S}_1 \cdot \mathbf{q} \right)  \left(\mathbf{S}_2 
\cdot \mathbf{q} \right)
	\right]\,,
\label{U-em}
\end{eqnarray}
where $\Pi_{ss}$ is the plasma correction to the space-space component
of  the photon propagator. 

If, as above, we assume that the wave function of $W$-bosons is space
independent and average this potential over space, we obtain
the following expression for the spin-spin part of the energy shift:
\begin{eqnarray} 
\delta E =\int \frac{d^3 r}{V} U^{(spin)}_{em}  ({\bf r}) =
- \frac{e^2 \rho^2}{V m_W^2} 
\int \frac{d^3 q}{(2\pi)^3} \delta^{(3)} \! ({\bf q}) \,\,
\frac{ q^2 ({\bf S}_1\cdot {\bf S}_2)  - ({\bf  q \cdot S}_1) ({\bf q \cdot S}_2)}
{q^2 + \Pi_{ss} ({\bf q})}
\label{delta-of-q}
\end{eqnarray}
Clearly $\delta E$ vanishes if $\Pi_{ss}$ is non-zero at $q = 0$.
Of course, this is an unphysical conclusion, because the integration
over $r$ should be done with some finite upper limit, $r_{max}=l$, 
presumably equal to the average distance between the $W$ bosons.
So instead of the delta-function, $\delta^{(3)} ({\bf q})$, we obtain:
\begin{eqnarray}
\int_0^l d^3 r \exp (i{\bf qr}) = \frac{4\pi}{q^3}  \left[ \sin (ql) - 
ql \cos (ql) \right].
\label{int-dr}
\end{eqnarray}
The energy shift is given by the expression:
\begin{eqnarray}
\delta E =
- 4 \pi \frac{e^2 \rho^2}{V m_W^2}
S_1^i S_2^j
\int \frac{d^3q}{(2\pi)^3} \,
\frac{\left[\sin(ql) - ql \cos(ql)\right] \left[q^2 \delta_{ij} - q^i q^j \right]}
{q^3\left[ q^2 + \Pi_{ss}({q}) \right]},
\label{delta-E}
\end{eqnarray}
where $V = 4 \pi l^3 /3$.

When we average over an angle independent wave function, e.g. S-wave for the condensate, the non-vanishing part of the integral in Eq. (\ref{delta-E}) is proportional to the Kronecker delta, hence:
	\begin{eqnarray}
	\delta E 
	= S_1^i S_2^j A \, \delta^{ij},
	\end{eqnarray}
where the coefficient $A$ can be calculated by taking trace of Eq. (\ref{delta-E}):
	\begin{eqnarray}
	Tr (A \delta^{ij}) = 3A = 
	- 8 \pi \frac{e^2 \rho^2}{m_W^2} 
	\int \frac{d^3q}{(2\pi)^3} \,
	\frac{\left[q\sin(ql) - q^2l \cos(ql)\right] }
	{q^2\left[ q^2 + \Pi_{ss}({q}) \right]}.
	\end{eqnarray}
Hence the energy shift of a pair of $W$-bosons in S-wave state
due to the spin-spin interaction is:
	\begin{eqnarray}
	\delta E = - \left(\mathbf{S}_1 \cdot \mathbf{S}_2 \right) 
	 \frac{8 \pi e^2 \rho^2}{3 V m_W^2} 
	\int \frac{d^3q}{(2\pi)^3} \,
	\frac{\left[\sin(ql) - ql \cos(ql)\right] }
	{q\left[ q^2 + \Pi_{ss}({q}) \right]}\,.
	\end{eqnarray}
Introducing the new integration variable $x= ql$, we can rewrite it as:
\begin{eqnarray}
\delta E
= - ({\bf S}_1 \cdot {\bf S}_2) \frac{4 e^2 \rho^2}{3\pi V m_W^2}
\int_0^\infty \frac{dx }
{x^2 + l^2\Pi_{ss}(x/l)} \left[x\,\sin x + l^2\Pi_{ss}(x/l) \cos x\right]\,,
\end{eqnarray}
We used here the usual regularization of divergent integrals: 
$\exp(\pm i ql) \rightarrow \exp(\pm i ql - \epsilon q)$ with
$\epsilon \rightarrow 0$. 
With such regularization $\int_0^\infty dx \cos(x) = 0$.

Evidently, if $\Pi_{ss} = 0$, we obtain the same result as that found
above.  In fact the necessary condition for
obtaining the ``unscreened'' result is $l^2 \Pi_{ss}(x/l) \ll 1$, but
for a large $l^2 \Pi_{ss}$ the electromagnetic 
part of the spin-spin interaction can be suppressed enough to change the
ferromagnetic behavior into the antiferromagnetic one. This might take place
at high temperatures above the EW phase transition when the Higgs
condensate is destroyed and the masses of $W$ and $Z$ appear only as
a result of temperature and density corrections and thus are
relatively small. The quantitative statement depends upon the
modification of the space-space part of the photon propagator in
presence of the Bose condensate of charged $W$. As far as we know,  this
modification has not yet been found. 

If $W$-bosons make a ferromagnetic state, the primeval plasma, where such
bosons  condensed (possibly due to a  large cosmological lepton
asymmetry), could be spontaneously magnetized, as it happens
in usual ferromagnets. The typical size of the magnetic
domains is determined by the cosmological horizon at the moment of the
condensate evaporation. The latter takes place when the neutrino chemical
potential, which scales as temperature in the course of cosmological
cooling down, becomes smaller than the $W$ mass at this temperature.

The large scale magnetic field, produced 
by the ferromagnetism of $W$-bosons,
might survive after the decay of the condensate due to the 
conservation of the magnetic flux in the primeval
plasma because of its high electric conductivity.
Such magnetic fields, which were uniform at macroscopically 
large scales at formation, may be the seeds for the
observed galactic or intergalactic magnetic fields at astronomically
large scales. Evidently the characteristic size of magnetic domains
at production is much smaller than the galactic size, even with an
account of the cosmological stretching out. Nevertheless, magnetic fields
homogeneous at astronomical scales may be created by chaotic
reconnection (Brownian motion) of the magnetic field lines at much 
smaller scales but by an expense of the field amplitude. Such a mechanism
could be a competing alternative among many other attempts to solve
the mystery of the generation of large scale magnetic fields in
cosmology, for a review see ref.~\cite{magn-rev}.


\begin{thebibliography}{9}
\bibitem{landau-9}
E.M. Lifshitz and L.P Pitaevskii, {\it Landau and Lifshitz, Course of Theoretical 
Physics, Volume 10 - Physical Kinetics}, Elsevier, 1981.

\bibitem{kapusta}
J.~I.~Kapusta, C.~Gale,  
{\it Finite temperature field theory: Principles and Applications},
Cambridge Monographs on Mathematical Physics, 2006.

\bibitem{fradkin}
I.~A.~Akhiezer and C.~V.~Peletminsky, {\em ZhETF} {\bf 38}, 1829 (1960); Sov. Phys. JETP, 
{\bf 11}, 1316 (1960);\\ 
E.~S.~Fradkin, {\em Proc. Lebedev Inst.} {\bf 29}, 7 (1965).

\bibitem{friedel}
J. Friedel, {\em Phil. Mag.} {\bf 43}, 153 (1952); Nuovo Cim. {\bf 7}, 287 (1958), suppl. 2;\\
J.S. Langer, S.H. Vosko, {\em J. Phys. Chem. Solids} {\bf 12}, 196 (1960);\\
A. Fetter, J. Walecka, {\it Quantum Theory of Many-Particle Systems}, McGraw-Hill, 
San Francisco, 1971.

\bibitem{gaba}
 G. Gabadadze,  R.A. Rosen,
 {\em JCAP} {\bf 0810} (2008) 030
  [arXiv:0806.3692 [astro-ph]].
\\
 G.~Gabadadze and R.~A.~Rosen,
  {\em JCAP} {\bf 0902} (2009) 016
  [arXiv:0811.4423 [hep-th]].
\\
 G.~Gabadadze and R.~A.~Rosen,
  {\em JCAP} {\bf 1004} (2010) 028
  [arXiv:0912.5270 [hep-ph]].


\bibitem{dlp}
 A.D. Dolgov, A. Lepidi, G.Piccinelli,
  {\em JCAP} {\bf 0902} (2009) 027
  [arXiv:0811.4406 [hep-th]].
\\
 A.D. Dolgov, A. Lepidi, G. Piccinelli,
  {\em Phys.\ Rev.  D} {\bf 80} (2009) 125009
  [arXiv:0905.4422 [hep-ph]].

\bibitem{dlp-vec}
A.D. Dolgov, A. Lepidi, G. Piccinelli, {\em  JCAP}  {\bf 08} (2010) 031
[arXiv:1005.2702 [astro-ph.CO]].

\bibitem{gaba-vec}
L. Berezhiani, G. Gabadadze,  D. Pirtskhalava,
 {\em  JHEP} {\bf 1004} (2010) 122
  [arXiv:1003.0865 [hep-ph]].

\bibitem{Pethick_book}
  C. J. Pethick and H. Smith,
  {\it Bose-Einstein Condensation in Dilute Gases}, Cambridge University Press, 2002.

\bibitem{linde-screen}
A. D. Linde, {\em Rep.  Prog.  Phys.} {\bf 42} (1979) 389;\\
A. D. Linde, {\em Phys. Lett.} 96B (1980) 293;\\
D. J. Gross, R. Pisarski, L. Yaffe,  {\em Rev. Mod. Phys.} {\bf 53} (1981) 43.

\bibitem{magn-rev}
  D.~Grasso and H.~R.~Rubinstein,
  Phys.\ Rept.\  {\bf 348} (2001) 163
  [arXiv:astro-ph/0009061];
\\
A.D. Dolgov, In Gurzadyan, V.G. (ed.) et al.: From integrable models to gauge theories, 143-154. 
e-Print: hep-ph/0110293;\\
A.D. Dolgov,
Talk given at 17th Les Rencontres de Physique de la Vallee d'Aoste: Results and Perspectives in Particle Physics, La Thuile, Aosta Valley, Italy, 9-15 Mar 2003, e-Print: astro-ph/0306443;\\
  M.~Giovannini,
  Int.\ J.\ Mod.\ Phys.\  D {\bf 13} (2004) 391
  [arXiv:astro-ph/0312614].


\end{thebibliography}
\end{document}